\documentclass[12pt, a4paper]{article}
\usepackage{cite}
\usepackage{amsmath,amssymb}
\input{colordvi.tex}
\usepackage{comment}
\usepackage{bm}
\usepackage{url}
\usepackage{ifpdf}
\ifpdf        
  \usepackage{graphicx, hyperref, xcolor}     
\else     
  \usepackage[dvipdfmx]{graphicx, hyperref, xcolor}     
 \fi

\definecolor{rossoferrari}{HTML}{D9073D}
\definecolor{mediumblue}{HTML}{0000CD}
\hypersetup{
setpagesize=false,
bookmarksnumbered=true,%
bookmarksopen=true,%
colorlinks=true,%
linkcolor=rossoferrari,
urlcolor=mediumblue,
citecolor=mediumblue,
}

\begin{document}

\begin{titlepage}

\begin{center}

\hfill UT-19-11\\

\vskip .75in

{\Large \bf 
A Note on Gravitational Particle Production  \\ \vspace{2mm} in Supergravity
}

\vskip .75in

{\large
Kazunori Nakayama$^{(a,b)}$
}

\vskip 0.25in

$^{(a)}${\em Department of Physics, Faculty of Science,\\
The University of Tokyo,  Bunkyo-ku, Tokyo 113-0033, Japan}\\[.3em]
$^{(b)}${\em Kavli IPMU (WPI), The University of Tokyo,  Kashiwa, Chiba 277-8583, Japan}

\end{center}
\vskip .5in

\begin{abstract}

It is pointed out that the gravitational particle production rate of a scalar component of a chiral superfield in supergravity with minimal K\"ahler potential can be significantly suppressed compared with a minimal scalar field in non-supersymmetric Einstein gravity.
This suppression is avoided for some choice of the inflaton sector and also for non-minimal K\"ahler potential of the chiral superfield. 

\end{abstract}

\end{titlepage}


\renewcommand{\thepage}{\arabic{page}}
\setcounter{page}{1}
\renewcommand{\thefootnote}{\#\arabic{footnote}}
\setcounter{footnote}{0}

\newpage

\section{Introduction}
\label{sec:Intro}

Inflationary universe has become the standard cosmological scenario, although it is still difficult to pin down the model of inflation. 
One of the key ingredients to understand physics of inflation is reheating. The reheating is the epoch followed by inflation, during which various particles are somehow excited and the universe becomes filled by high-temperature plasma.
How the reheating proceeds depends on the detailed properties of the inflaton and it is highly model dependent~\cite{Dolgov:1989us,Traschen:1990sw,Shtanov:1994ce,Kofman:1997yn}. However, there is a ubiquitous process of particle production which does not much depend on the details of inflation models: gravitational particle production~\cite{Parker:1969au,Birrell:1982ix}.

All the fields feel the background expansion of the universe unless they are conformally coupled to gravity and as a result a vacuum state cannot remain exactly the same throughout the cosmological history. It may be viewed as a particle production due to the gravitational effect.
Gravitational particle production is often efficient in the early universe, in particular around the epoch of transition from inflation to the matter- or radiation-dominated universe~\cite{Ford:1986sy}. 
There are several possible phenomenological consequences: the total radiation may be created in this way in some specific inflation model~\cite{Peebles:1998qn}, a stable particle which has only the gravitational interaction can be a dominant component of dark matter~\cite{Chung:1998zb,Chung:2001cb,Chung:2011ck}, some cosmologically harmful particles such as moduli can be created~\cite{Giudice:1999yt}, and so on. 
Recently it was proposed that the gravitational particle production is also efficient at the reheating stage where the inflaton oscillates rapidly~\cite{Ema:2015dka,Ema:2016hlw} and it also has impacts on the production of dark matter~\cite{Ema:2018ucl,Chung:2018ayg,Ema:2019yrd}.
See also Refs.~\cite{Bassett:1997az,Tsujikawa:1999jh,Watanabe:2007tf,Markkanen:2015xuw,Addazi:2017ulg,Hashiba:2018iff,Li:2019ves} for related topics and applications.

In this paper we explore gravitational particle production in supergravity.
Supersymmetry (SUSY) solves or relaxes the gauge hierarchy problem and it also leads to the gauge coupling unification at the scale of grand unification~\cite{Martin:1997ns}. Thus the SUSY extension of the standard model is motivated and global SUSY is naturally extended to supergravity.
In the minimal SUSY standard model (MSSM), the lightest SUSY particle (LSP) is a candidate of dark matter if the R-parity is conserved.
However, R-parity needs not be exact and in such a case we may need other candidate of dark matter. As the simplest possibility, one can introduce a gauge singlet chiral superfield $\chi$ with a superpotential $W=m_\chi \chi^2/2$, whose stability may be guaranteed by $Z_2$ symmetry under which $\chi$ transforms as $\chi \to -\chi$.
In this kind of setup, we need to evaluate the rate of gravitational particle production in the framework of supergravity.
Besides dark matter, gravitational production of dangerous long-lived particles such as Polonyi/moduli can be cosmologically relevant and hence it is important to know the rate of gravitational particle production.
As shown in detail later, Planck-suppressed interactions that necessarily appear in supergravity can significantly affect the production rate.

The spontaneous inflaton decay induced by supergravity effect was extensively studied in a series of works~\cite{Endo:2006qk,Endo:2006nj,Endo:2007sz}, which may also be viewed as gravitational particle production in supergravity. We rather focus on the situation where such decay processes are forbidden or inefficient, which is actually the case depending on the property of the inflaton.

In Sec.~\ref{sec:GP} we briefly review the gravitational particle production in non-SUSY Einstein gravity.
In Sec.~\ref{sec:sugra} the gravitational particle production rate in supergravity is studied.
In particular, we find that the production rate of a scalar field with minimal K\"ahler potential can be suppressed compared with the non-SUSY case.
The case of more general classes of K\"ahler potential and several additional features of gravitational particle production in supergravity are discussed in Sec.~\ref{sec:ext}.
Some phenomenological implications are discussed in Sec.~\ref{sec:dis}.

\section{Gravitational particle production in non-SUSY theory}
\label{sec:GP}

First let us briefly review the gravitational particle production in the non-SUSY case with Einstein gravity.
We introduce a scalar field $\chi$, which is minimally coupled to gravity:
\begin{align}
	S = \int d^4x \sqrt{-g}\left(\frac{M_P^2}{2}R - \frac{1}{2}g^{\mu\nu}\partial_\mu\chi \partial_\nu\chi - \frac{1}{2} m_\chi^2\chi^2 - \mathcal L_{\phi} \right),
\end{align}
where $M_P$ is the reduced Planck scale and $R$ the Ricci curvature, $\phi$ denotes the canonical inflaton field and $\mathcal L_\phi$ is the inflaton Lagrangian, whose concrete form is not specified.
We assume the Friedmann-Robertson-Walker universe where the line element is given by $ds^2=-dt^2+a^2(t) d\vec x^2$ with $a(t)$ being the cosmic scale factor. 
To discuss the gravitational particle production of $\chi$, it is convenient to redefine the canonical field $\widetilde\chi \equiv a \chi$ and use the conformal time $d\tau = dt /a(t)$ and rewrite the action as
\begin{align}
	S_\chi = \int d\tau d^3x \frac{1}{2}\left[  \widetilde\chi'^2 -(\nabla\widetilde\chi)^2- \widetilde m_\chi^2 \widetilde\chi^2 \right],
	\label{Schi_NSUSY}
\end{align}
where the prime denotes the derivative with respect to conformal time and 
\begin{align}
	\widetilde m_\chi^2 \equiv a^2 m_\chi^2 - \frac{a''}{a} = a^2\left(m_\chi^2 - \frac{R}{6}\right)
	= a^2 \left( m_\chi^2 -2H^2 + \frac{\dot\phi^2}{2M_P^2} \right),
	\label{mchi_NSUSY}
\end{align}
where $H=\dot a/a$ denotes the Hubble parameter and we have used the Friedmann equation in the last equality:
\begin{align}
	H^2 = \left(\frac{\dot a}{a}\right)^2= \frac{1}{3M_P^2}\left(\frac{1}{2}\dot\phi^2 + V_\phi\right),
\end{align}
with $V_\phi$ being the inflaton potential. Eq.~(\ref{mchi_NSUSY}) may show that the minimal scalar field $\chi$ ``feels'' the time dependence of the Ricci curvature $R$.

Particle production phenomenon is related to the (rapid) time-dependence of the effective mass $\widetilde m_\chi^2$. It is clearly seen that the last term in the most right hand side of the expression (\ref{mchi_NSUSY}) is highly time-dependent due to the inflaton dynamics, while the time dependence of the first and second terms are rather mild compared with the last term.

Below we estimate the gravitational production of $\chi$ particle in the two regimes: the transition era from de Sitter to radiation or matter dominated universe, and the reheating era during which the universe is dominated by the inflaton coherent oscillation.
We focus on the case of $m_\chi^2 \gtrsim 2H_{\rm inf}^2$ where $H_{\rm inf}$ denotes the Hubble scale during inflation, since otherwise superhorizon quantum fluctuations are amplified during inflation and $\chi$ obtains a nearly constant field value in the universe, which would results in the coherent oscillation of $\chi$ and it is often the dominant contribution to the $\chi$ abundance.
On the other hand, if $\chi$ is heavier than any other mass scale appearing in the background dynamics, no particle production would be expected. The inflaton mass $m_\phi$ around its potential minimum (or the inverse time scale of the inflaton oscillation time period) may be the largest mass scale. Thus we are interested in the case of $H_{\rm inf} \lesssim m_\chi \lesssim m_\phi$.

\subsection{Production at the transition}  \label{sec:slow_NSUSY}

The particle production rate can be estimated by the standard procedure and readers are referred to e.g. Refs.~\cite{Birrell:1982ix,Kofman:1997yn,Ema:2016hlw,Ema:2018ucl} for detailed formulation. 
Here we just summarize the result. The phase space density of $\chi$ is given by $f_\chi(k,\tau) = |\beta_k(\tau)|^2$ where
\begin{align}
	\beta_k(\tau) \simeq \int_{\tau_i}^{\tau} \frac{\omega_k'}{2\omega_k}\exp\left(-2i \int_{\tau_i}^{\tau} \omega_k d\tau' \right)d\tau,~~~~~~
	\omega_k^2 \equiv k^2 + \widetilde m_\chi^2.
	\label{betak}
\end{align}
The number density is then given by
\begin{align}
	a^3(\tau) n_\chi(\tau) = \int \frac{d^3k}{(2\pi)^3} f_\chi(k,\tau).
\end{align}
To evaluate (\ref{betak}), the time dependence of the integrand $\omega_k'/(2\omega_k)$ should be carefully treated.
It is important to notice that there are two distinct time scales: the Hubble time scale just caused by the universe expansion and the inflaton oscillation time scale in the reheating period.
Since the Hubble time scale is much slower than the inflaton oscillation time scale, we call the former as ``slow'' contribution and the latter as ``fast'' contribution.

Let us first estimate the slow contribution. Since we are interested in the case of $m_\chi \gtrsim H_{\rm inf}$ as mentioned above, the integrand may be approximated as
\begin{align}
	\frac{\omega_k'}{2\omega_k} \simeq \frac{a^2 \mathcal H m_\chi^2}{2(k^2+a^2m_\chi^2)}.
\end{align}
The integration (\ref{betak}) is dominated around the epoch when the integrand changes its slope as a function of $\tau$.\footnote{
	For fixed wavenumber $k$, the integration is dominated either at the epoch of transition from de Sitter to matter-dominated universe or the horizon crossing $k \sim aH$ or when it becomes non-relativistic $k=am_\chi$.
}
The calculation is the same as the case of a scalar field with conformal coupling to gravity, and the resulting number density is given by~\cite{Ema:2018ucl}
\begin{align}
	n_\chi^{\rm (slow)} (\tau)\simeq \mathcal A\,m_\chi^3 e^{-c m_\chi / H_{\rm end}} \left( \frac{a(\tau_{\rm end})}{a(\tau)} \right)^3,
\end{align}
where $\mathcal A\sim 10^{-3}$ and $c\sim 5$ are numerical constants and $\tau_{\rm end}$ denotes the conformal time at the end of inflation and $H_{\rm end}$ denotes the Hubble scale around the end of inflation. It is noticed that $H_{\rm inf}$ and $H_{\rm end}$ are the same order in general.

\subsection{Production during inflaton coherent oscillation}  \label{sec:fast_NSUSY}

Next let us consider the period of inflaton coherent oscillation, during which the inflaton potential may be approximated as $V(\phi) \simeq m_\phi^2\varphi^2/2$ where we have defined $\varphi \equiv \phi - v_\phi$ with $v_\phi$ denoting the inflaton field value at the potential minimum.
As pointed out in Ref.~\cite{Ema:2015dka}, not only the last term in (\ref{mchi_NSUSY}) but also the first and second terms contain rapidly oscillating part, because one can approximate
\begin{align}
	a(t) \simeq \left<a(t)\right>\left(1-\frac{\varphi^2-\left<\varphi^2\right>}{8M_P^2} \right),~~~~~~
	H(t) \simeq \left< H(t) \right>-\frac{\varphi \dot \varphi}{4M_P^2},
	\label{H_osc}
\end{align}
where the bracket shows the time average over the inflaton oscillation period. Therefore, the oscillating part of the effective mass (\ref{mchi_NSUSY}) may be estimated as
\begin{align}
	\left[ \widetilde m_\chi^2 \right]_{\rm osc} \simeq  \left<a^2\right>
	 \left( -(m_\chi^2-2 \left<H\right>^2)\frac{\varphi^2}{4M_P^2}+ \left<H\right> \frac{\varphi\dot\varphi}{M_P^2} + \frac{\dot\varphi^2}{2M_P^2}  \right).
	 \label{m_osc_NSUSY}
\end{align}
During the inflaton oscillation regime we have $m_\phi \gg H$ and also we are interested in the case $m_\chi < m_\phi$ because otherwise there is no $\chi$ particle production. Thus the last term is the dominant one.

Particle production due to the oscillating mass term was studied in detail in the context of reheating/preheating~\cite{Dolgov:1989us,Traschen:1990sw,Shtanov:1994ce,Kofman:1997yn}. Analytic and numerical studies to evaluate (\ref{betak}) with such an oscillating function due to gravity effect have been done in Refs.~\cite{Ema:2018ucl,Chung:2018ayg}.
In this case, the eventual number density of $\chi$ particle is dominated by those created earlier epoch and estimated as
\begin{align}
	n_\chi^{\rm (fast)}(\tau) \simeq \mathcal C H_{\rm end}^3 \left(\frac{a(\tau_{\rm end})}{a(\tau)}\right)^3,
	\label{nfast_NSUSY}
\end{align}
where $\mathcal C = 3/(128\pi)$ is a numerical constant. Numerically we find $\mathcal C \simeq 7\times 10^{-3}$ which agrees well with this analytic expression.

\section{Gravitational particle production in supergravity}
\label{sec:sugra}

Let us consider a model with minimal K\"ahler potential for $\chi$ and the following simple superpotential:
\begin{align}
	&K = |\chi|^2 + |X|^2 -\frac{1}{2}(\phi - \phi^\dagger)^2, \label{K_min}\\
	&W = \frac{1}{2} m_\chi \chi^2 + X g(\phi) + W_0.  \label{W_sugra}
\end{align}
Here $\chi$ is a chiral superfield with mass $m_\chi$. The inflaton sector consists of $X$ and $\phi$ as usual with $X$ being often called as a stabilizer field. We have taken a shift-symmetric K\"ahler potential for $\phi$~\cite{Kawasaki:2000yn}, which ensures that ${\rm Im}(\phi)$ is stabilized during inflation (by adding some higher order terms like $K \sim |X|^2(\phi-\phi^\dagger)^2/M_P^2$, if necessary) and $\sqrt{2} {\rm Re}(\phi)$ is regarded as a physical inflaton field. 
Thus the physical inflaton field does not show up in the K\"ahler potential and it simplifies the analysis below.\footnote{
	To be precise, the reality of $g(\phi)$ is required so that the inflaton direction does not change between inflation and the reheating stage.
}
The constant term is related to the present gravitino mass $m_{3/2}$ through $W_0 = m_{3/2} M_P^2$. Without loss of generality, we can take $m_\chi$ and $W_0$ real and positive.
The scalar potential is given by
\begin{align}
	V = e^{K/M_P^2} \left[ K^{i\bar j}(D_i W)(D_{\bar j}\overline W) - \frac{3|W|^2}{M_P^2} \right],
\end{align}
where $D_iW \equiv \partial_i W + W\partial_i K / M_P^2$ with $i=\chi, X,\phi$.
Here and in what follows we use the same character for a chiral superfield and its scalar component.
For a while let us take $X=0$. See Sec.~\ref{sec:gravitino} for effects of $X$ dynamics. The action of the scalar field $\chi$, up to the second order in $\chi$, is given as
\begin{align}
	S = \int d^4x \sqrt{-g} \left( - |\partial \chi |^2 - V_\chi \right),
\end{align}
where\footnote{
	Note that there should be additional SUSY breaking sector whose potential cancels the supergravity contribution $-3|W_0|^2/M_P^2$ in the vacuum.
}
\begin{align}
	V_\chi \simeq \left(m_\chi^2+m_{3/2}^2+ \frac{V_{\phi}}{M_P^2}\right)|\chi|^2 -\frac{1}{2}m_\chi m_{3/2} (\chi^2 + \chi^{*2}),
	\label{Vchi_sugra}
\end{align}
where $V_\phi \simeq |g(\phi)|^2$ denotes the inflaton potential energy. 
As is evident, the scalar field mass receives supergravity corrections. Substituting $\chi=(\chi_R+i \chi_I)/\sqrt{2}$, it is found that the real and imaginary part do not mix with each other and have masses of
\begin{align}
	&m_-^2 = m_{\chi,R}^2+ \frac{V_\phi}{M_P^2},~~~~~~m_{\chi,R}^2 \equiv m_\chi^2+m_{3/2}^2 -\frac{1}{2}m_\chi m_{3/2},\\
	&m_+^2 =m_{\chi,I}^2+ \frac{V_\phi}{M_P^2},~~~~~~m_{\chi,I}^2 \equiv m_\chi^2+m_{3/2}^2 +\frac{1}{2}m_\chi m_{3/2},
	\label{mplus}
\end{align}
respectively.
In terms of the rescaled field $\widetilde\chi_j \equiv a\chi_j$ $(j=R,I)$, as done in (\ref{Schi_NSUSY}), we can rewrite the action as
\begin{align}
	S_\chi = \int d\tau d^3x \sum_{j=R,I} \frac{1}{2}\left[  \widetilde\chi'^2_j -(\nabla\widetilde\chi_j)^2- \widetilde m_{\chi,j}^2 \widetilde\chi_j^2 \right],
	\label{Schi_sugra}
\end{align}
where
\begin{align}
	 \widetilde m_{\chi,j}^2 \equiv a^2\left(m_{\chi,j}^2+ \frac{V_\phi}{M_P^2}\right) - \frac{a''}{a}
	 = a^2\left(m_{\chi,j}^2+ H^2\right).
	 \label{mchi_sugra}
\end{align}
Remarkably, thanks to the supergravity correction $V_\phi/M_P^2$, the effective mass of the minimal scalar field becomes dependent only on the Hubble parameter $H$. This is in contrast to the non-SUSY case (\ref{mchi_NSUSY}) where the effective mass depends on the Ricci curvature $R$.

The difference between these two is significant when we consider gravitational particle production due to the inflaton coherent oscillation,
since the Ricci curvature $R$ is a violently oscillating function while the Hubble parameter $H$ is approximately an adiabatic function.
This is because the Hubble parameter squared is proportional to the energy density through the Friedmann equation and the energy density is approximately a conserved quantity.  Below we estimate the production rate of $\chi$. For simplicity we assume $m_{3/2} \ll m_\chi$ hereafter.

\subsection{Production at the transition}  \label{sec:slow_sugra}

In the present case we can consider the both cases of $m_\chi \ll H_{\rm inf}$ and $m_\chi \gg H_{\rm inf}$ since the long wave fluctuations do not develop during inflation even in the limit $m_\chi^2\to 0$ because of the supergravity correction to the mass term.
For $m_\chi \gtrsim H_{\rm end}$, the integrand is approximated by
\begin{align}
	\frac{\omega_k'}{2\omega_k} \sim \frac{a^2 \mathcal H m_\chi^2}{2(k^2+a^2m_\chi^2)}.
\end{align}
It is similar to the case of the scalar field with conformal coupling to gravity, and the resulting number density is given by
\begin{align}
	n_\chi^{\rm (slow)} (\tau)\simeq \mathcal A\,m_\chi^3 e^{-c m_\chi / H_{\rm end}} \left( \frac{a(\tau_{\rm end})}{a(\tau)} \right)^3.
\end{align}
For $m_\chi \lesssim H_{\rm end}$, on the other hand, the production rate becomes insensitive to the mass $m_\chi$. The result is
\begin{align}
	n_\chi^{\rm (slow)} (\tau)\simeq \mathcal A'\, H_{\rm end}^3 \left( \frac{a(\tau_{\rm end})}{a(\tau)} \right)^3,
\end{align}
where we find $\mathcal A' \sim 10^{-4}$ by numerical calculation.
Thus the slow contribution can be a similar order to the non-SUSY case studied in Sec.~\ref{sec:slow_NSUSY}.
One should note, however, that the slow contribution is exponentially suppressed as $e^{-cm_\chi/H_{\rm end}}$ for $m_\chi \gtrsim H_{\rm end}$.
In such a case, it is only the fast contribution that can create $\chi$ particles through gravitational interaction.

\subsection{Production during inflaton coherent oscillation} \label{sec:fast_sugra}

Now let us estimate the gravitational production of $\chi$ particle through the fast contribution in supergravity. 
During the inflaton coherent oscillation, what is important for the particle production is the oscillating part of the effective mass (\ref{mchi_sugra}).
Using Eq.~(\ref{H_osc}), it is given by
\begin{align}
	 \left[ \widetilde m_{\chi,j}^2 \right]_{\rm osc} \simeq 
	 \left<a^2\right> \left( -\frac{m_{\chi}^2 \varphi^2}{4M_P^2}  -\left< H \right> \frac{\varphi\dot\varphi}{2M_P^2}\right),
\end{align}
where we have defined $\varphi /\sqrt{2} \equiv {\rm Re}(\phi)-v_\phi$.
Compared with the non-SUSY case (\ref{m_osc_NSUSY}), the typical magnitude of the oscillating effective mass is suppressed by the factor $\epsilon$:
\begin{align}
	\epsilon \equiv {\rm max}\left[ \frac{m_{\chi}^2}{2m_\phi^2},~\frac{H}{m_\phi} \right].
	\label{eps_sugra}
\end{align}
Thus the rate of gravitational particle production in supergravity may suffer a huge suppression factor. 
Correspondingly, the number density of produced $\chi$ particle is given by
\begin{align}
	n_\chi^{\rm (fast)}(\tau) \simeq \mathcal C\, \epsilon^2\,H_{\rm end}^3 \left(\frac{a(\tau_{\rm end})}{a(\tau)}\right)^3.
	\label{nfast_sugra}
\end{align}
Numerically we find $\mathcal C \sim 0.04$.
In low scale inflation models we generically have $H_{\rm end} (\sim H_{\rm inf}) \ll m_\phi$.
Therefore, unless $m_\chi \sim m_\phi$ accidentally, the particle production rate may be orders of magnitude suppressed compared with the case of minimal scalar in non-SUSY theory.
Viewed as $\phi\phi\to\chi\chi$ scattering process, this suppression might be understood as a cancellation of the scattering amplitude between terms like $m_\phi^2|\phi|^2 |\chi|^2/ M_P^2$ and $|\phi|^2 |\partial_\mu\chi|^2/M_P^2$.

\begin{figure}
\begin{center}
\begin{tabular}{cc}
\includegraphics[scale=1.3]{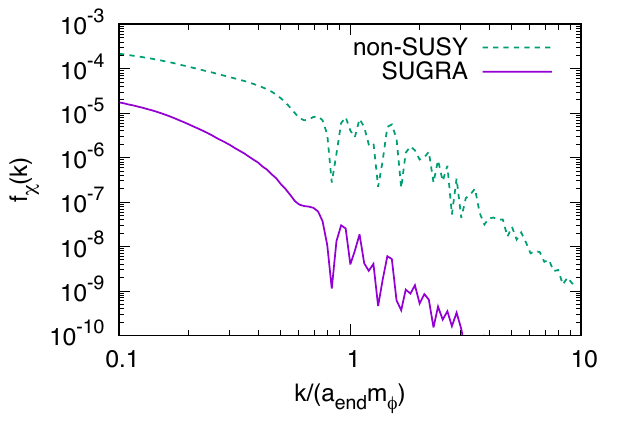}
\includegraphics[scale=1.3]{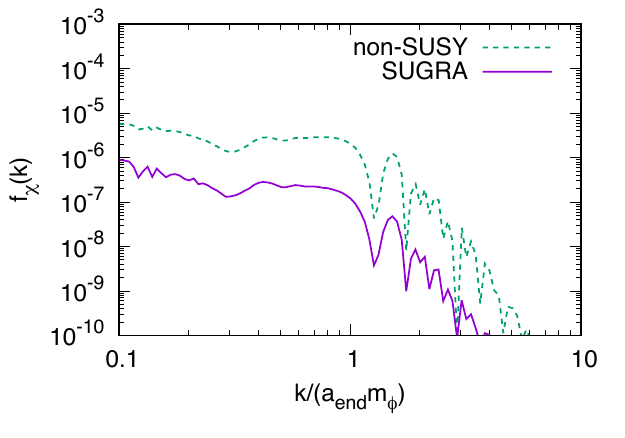}
\end{tabular}
\end{center}
\caption{
	The phase space density of $\chi$ particle well after inflation due to the gravitational particle production in the new inflation model (\ref{V_new}).
	We have taken $m_\chi=0.2 m_\phi$ (left) and $m_\chi=m_\phi$ (right).
}
\label{fig:fk}
\end{figure}

We have performed numerical calculations to evaluate the phase space density $f_\chi(k)$.
To do so, as a concrete example we adopt a new inflation model whose potential is described by
\begin{align}
	V_\phi = M^4 \left[ 1- \left(\frac{\phi}{v_\phi}\right)^{n}\right]^2.  \label{V_new}
\end{align}
We have taken $n=6$ since $n\geq 6$ is favored from the Planck observation. In SUSY, this potential is realized by assuming
\begin{align}
	g(\phi) = M^2\left[ 1-  \left(\frac{\sqrt{2}\phi}{v_\phi}\right)^{n} \right],
\end{align}
in the superpotential (\ref{W_sugra})~\cite{Asaka:1999jb,Senoguz:2004ky}.
We take both $M$ and $v_\phi$ real and positive, although it is not guaranteed in general for the shift-symmetric K\"ahler potential (\ref{K_min}) unless additional CP symmetry in the inflaton sector is introduced.\footnote{
	The SUSY new inflation model suffers from the $\eta$ problem~\cite{Copeland:1994vg} if the minimal K\"ahler potential $K=|\phi|^2$ would be taken. For successful inflation, a tuning is required for the coefficient of the additional term $K = |\phi|^2|X|^2/M_P^2$. This problem does not exist for the shift-symmetric K\"ahler potential.
}
The inflaton mass around the potential minimum is given by $m_\phi = \sqrt{2}n M^2/v_\phi$, which is much larger than the Hubble scale at the end of inflation $H_{\rm end} \simeq M^2/(\sqrt{3} M_P)$ for $v_\phi \ll M_P$.
We have solved the inflaton dynamics as well as Eq.~(\ref{betak}) for both the non-SUSY case and supergravity case. Note that both the slow and fast contributions are automatically included in the numerical study. Actually they are not exactly separable in a realistic setup.
Fig.~\ref{fig:fk} shows the resulting phase space distribution $f_\chi(k)=|\beta_k|^2$ well after the end of inflation.
We have taken $v_\phi=0.5M_P$, corresponding to $m_\phi \simeq 29 H_{\rm end}$, and $m_\chi=0.2 m_\phi$ (left) and $m_\chi=m_\phi$ (right).
The power law tail of the high momentum modes $k/(a(\tau_{\rm end})m_\phi) \gtrsim 1$ may be interpreted as a result of fast contribution, i.e., the gravitational production due to the inflaton coherent oscillation. Lower momentum modes are mixtures of slow and fast contributions.
It is seen that, for the light particle case $m_\chi = 0.2 m_\phi$ the fast contribution in supergravity is significantly suppressed compared with the non-SUSY case, while such a suppression is rather mild for the heavy case $m_\chi=m_\phi$, as expected from (\ref{nfast_sugra}).
Note that the total number density is dominated by the contribution around $k/(a(\tau_{\rm end})m_\phi) \sim 1$, which reproduces (\ref{nfast_sugra}).

\section{Extensions}
\label{sec:ext}

\subsection{Non-minimal K\"ahler potential}  \label{sec:nonmin}

We have shown that the rate of gravitational production, in particular the ``fast'' contribution, can be significantly suppressed in supergravity if the scalar field has a minimal K\"ahler potential like (\ref{K_min}).
The story changes if one allows to include non-minimal K\"ahler potential of $\chi$. For example, let us take
\begin{align}
	K = |\chi|^2 + |X|^2 -\frac{1}{2}(\phi-\phi^\dagger)^2 + c_X\frac{|X|^2 |\chi|^2}{M_P^2}.
	\label{K_nonmin}
\end{align}
Then we find that the resulting action of the canonical scalar $\widetilde\chi \equiv a h \chi$ is given by (\ref{Schi_sugra}) with
\begin{align}
	 \widetilde m_{\chi,j}^2 = \frac{a^2}{h^2}\left(m_{\chi,j}^2+ \frac{(1-c_X)V_\phi +c_X|\partial X|^2}{M_P^2} \right) - \frac{(ha)''}{ha},
\end{align}
where we have defined
\begin{align}
	h^2(X) \equiv 1 + \frac{c_X|X|^2}{M_P^2}.
\end{align}
Neglecting the dynamics of $X$, we can approximate it as
\begin{align}
	 \widetilde m_{\chi,j}^2 \simeq a^2 \left[ m_{\chi,j}^2 + H^2 - \frac{c_X V_\phi }{M_P^2} \right].
\end{align}
The term proportional to $c_X$ shows oscillatory behavior during the inflaton oscillation regime. 
The $\chi$ abundance from the fast contribution is given by
\begin{align}
	n_\chi^{\rm (fast)}(\tau) \simeq \mathcal C\, c_X^2\,H_{\rm end}^3 \left(\frac{a(\tau_{\rm end})}{a(\tau)}\right)^3.
	\label{nfast_nonmin}
\end{align}
It is a similar order to the non-SUSY case (\ref{nfast_NSUSY}) for $c_X \sim \mathcal O(1)$.
Thus the suppression of the gravitational production rate found in Sec.~\ref{sec:sugra} is not ubiquitous in supergravity but rather a special feature of the minimal K\"ahler potential.

\subsection{Effect of gravitino mass}  \label{sec:gravitino}

So far we have neglected the gravitino mass $m_{3/2}$ and correspondingly $X$ dynamics.
Actually it is easily stabilized at $X\sim 0$ during inflation by adding $K \sim -|X|^4/M_P^2$ in the K\"ahler potential.
Strictly speaking, however, some amount of $X$ oscillation is induced dynamically after inflation, and it leads to the oscillation of ``effective gravitino mass'' $m_{3/2}^{\rm (eff)} \equiv W/M_P^2 \sim X g(\phi) / M_P^2 + m_{3/2}$.
As noted in Ref.~\cite{Ema:2016oxl}, the overall magnitude of $m_{3/2}^{\rm (eff)}$ is at most $m_{3/2}$.
Thus the leading correction to the $\chi$ potential (\ref{Vchi_sugra}) is
\begin{align}
	\delta V_\chi \simeq - \frac{m_{3/2}(Xg(\phi) + {\rm h.c.})}{M_P^2} |\chi|^2.
\end{align}
By using the estimate of $X$ amplitude presented in Sec.~4 of Ref.~\cite{Ema:2016oxl}, we obtain the following extra contribution to $\epsilon$ defined in (\ref{eps_sugra}) as
\begin{align}
	\delta \epsilon \sim \frac{m_{3/2}^2}{m_\phi(H+ m_{3/2})}.
\end{align}
Therefore, if $m_{3/2} \gtrsim H_{\rm end}$, it is possible that this term dominates the $\chi$ particle production.

\subsection{Fermion production} \label{sec:fermion}

In supergravity we necessarily have a fermionic partner of $\chi$, which we denote by $\psi$, and it is also produced gravitationally.
The fermion gravitational production was studied in detail in Refs.~\cite{Chung:2011ck,Ema:2019yrd}.\footnote{
	See also Refs.~\cite{Kallosh:2000ve,Nilles:2001fg,Ema:2016oxl} in the context of gravitino/inflatino preheating.
}
In the situation we are interested in, we do not need to take account of the mixing among $\chi$ and inflaton or the SUSY breaking field because of the $Z_2$ symmetry. Thus the fermion action is simply written as
\begin{align}
	S= \int d^4x \,e \left( -i e^\mu_a\,\overline{\psi} \overline\sigma^a D_\mu \psi  -\frac{1}{2} m_\psi (\psi\psi +{\rm h.c.}) \right),
\end{align}
where $e^\mu_a$ denotes the vierbein and $e\equiv \det(e^\mu_a)$ and $D_\mu$ denotes the covariant derivative.
The fermion mass is given by
\begin{align}
	m_\psi = e^{K/(2M_P^2)} m_\chi.   \label{mpsi}
\end{align}
Defining the canonical field $\widetilde\psi \equiv a^{3/2} \psi$, we have
\begin{align}
	S= \int d\tau d^3x \left( -i \delta^\mu_a\,\overline{\widetilde\psi} \overline\sigma^a \partial_\mu \widetilde\psi  -\frac{a}{2} m_\psi (\widetilde\psi\widetilde\psi +{\rm h.c.}) \right),
\end{align}
where $\partial_0$ should be understood as a derivative with respect to $\tau$.
It is conformal in the limit $m_\psi\to 0$ and hence the gravitational production rate is suppressed by powers of $m_\psi$.

The gravitational production is caused by the time dependence of the effective mass $am_\psi$.
For the shift-symmetric K\"ahler potential of the inflaton (\ref{K_min}), the physical inflaton field does not appear in the K\"ahler potential and we can just replace $K=0$ in (\ref{mpsi}). In this case we obtain the same result as a minimal non-SUSY case,
\begin{align}
	\left[ a m_\psi\right]_{\rm osc} \simeq -\left<a\right> m_\chi\, \frac{\varphi^2}{8M_P^2}.
\end{align}
Thus the production rate is also the same as the non-SUSY case studied in Refs.~\cite{Chung:2011ck,Ema:2019yrd}.
In particular, for $H_{\rm end} \ll m_\chi \lesssim m_\phi$, the fermion abundance is given by~\cite{Ema:2019yrd}
\begin{align}
	n_\psi(\tau) \simeq \mathcal C H_{\rm end}^3
	\left(\frac{m_\chi}{m_\phi}\right)^2 \left(\frac{a(\tau_{\rm end})}{a(\tau)}\right)^3.
	\label{npsi}
\end{align}
and it is larger than the scalar abundance (\ref{nfast_sugra}). If $m_\chi \ll H_{\rm end}$, on the other hand, the fermion abundance is suppressed by powers $m_\chi$ while the scalar abundance becomes insensitive to $m_\chi$ and hence the scalar abundance would dominate.

Note that the fermion production rate does not change much for non-minimal K\"ahler potential like (\ref{K_nonmin}) as far as $c_X \sim \mathcal O(1)$ while the scalar production is enhanced significantly as noticed in Sec.~\ref{sec:nonmin}. Thus in such a case the scalar production becomes dominant.

\subsection{Inflaton decay}  \label{sec:dec}

So far we have assumed shift-symmetric type K\"ahler potential for the inflaton $\phi$ as in (\ref{K_min}) and studied the production of $\chi$ and $\psi$ fields.
The story drastically changes if one adopts a minimal K\"ahler potential like $K = |\phi|^2$.
A significant difference between these two cases is that the physical inflaton does not appear in the K\"ahler potential in the former case while it does appear in the latter case.
For this minimal K\"ahler potential, the $\chi$ potential receives additional correction terms as 
\begin{align}
	\Delta V_\chi \sim \frac{m_\chi^2+2m_{3/2}^2}{M_P^2}|\phi|^2|\chi|^2 + \frac{m_\chi}{2M_P^2} \left[(\partial_\phi g)\,\phi^\dagger X^\dagger \chi^2 + {\rm h.c.}\right]
\end{align}
If $\phi$ oscillates around finite vacuum expectation value, one can expand $\phi = v_\phi + \delta\phi$ and $g(\phi) \simeq m_\phi \delta\phi$. Then one obtains terms like
\begin{align}
	V_\chi \supset \frac{(m_\chi^2+2m_{3/2}^2) v_\phi}{M_P^2}(\delta\phi + \delta \phi^*)|\chi|^2
	+  \frac{m_\chi m_\phi v_\phi}{2M_P^2} \left(X^\dagger \chi^2 + {\rm h.c.}\right),
\end{align}
which induce the decay of the inflaton into $\chi$ pair~\cite{Endo:2006qk,Endo:2006nj,Endo:2007sz}. Noting that $X$ and $\delta\phi$ are often maximally mixed with each other around the vacuum, the last term gives the dominant contribution to the inflaton decay.
Similarly, the inflaton-fermion interaction appears from (\ref{mpsi}) as
\begin{align}
	\mathcal L \simeq \frac{|\phi|^2}{4M_P^2} m_\chi \psi\psi +{\rm h.c.} \supset 
	\frac{m_\chi v_\phi (\delta\phi+\delta\phi^*)}{4M_P^2} \psi\psi + {\rm h.c.},
\end{align}
which also induces the inflaton decay into $\psi$ pair~\cite{Endo:2006qk,Endo:2006nj,Endo:2007sz}.
The inflaton decay rate into $\chi$ pair and $\psi$ pair are the same and estimated as\footnote{
	If we introduce K\"ahler potential as $K = -c_X' |\phi|^2 \chi^2/M_P^2 + {\rm h.c.}$, we would obtain the decay rate $\Gamma_{\rm dec} \sim c_X^{'2}v_\phi^2m_\phi^3/M_P^4$.
}
\begin{align}
	\Gamma_{\rm dec} \simeq \frac{1}{16\pi}\frac{m_\chi^2 v_\phi^2 m_\phi}{M_P^4}.  \label{Gamma_dec}
\end{align}

It may be compared with the gravitational production rate studied so far, presented in Eqs.~(\ref{nfast_sugra}), (\ref{nfast_nonmin}) and (\ref{npsi}), which may be reinterpreted as the ``annihilation rate'' of the inflaton~\cite{Ema:2015dka} as
\begin{align}
	\Gamma_{\rm ann} \simeq \mathcal C\frac{H^2 m_\phi}{3M_P^2} \times {\rm max}\left[ c_X^2,~\frac{H^2}{m_\phi^2},~\frac{m_\chi^2}{m_\phi^2} \right].
	\label{Gamma_ann}
\end{align}
At earlier time $\Gamma_{\rm ann}$ can be dominant or comparable but $\Gamma_{\rm dec}$ becomes more and more efficient at later time. 
Thus the contribution from the decay is expected to be dominant eventually.
These decay processes, however, vanish for $v_\phi=0$ or for the shift-symmetric K\"ahler potential as assumed in the most part of this paper.\footnote{
	The existence of shift-symmetric linear term in the K\"ahler potential $K \propto (\phi-\phi^\dagger)$ again causes the inflaton decay into $\chi$ and $\psi$, although it is forbidden by imposing $Z_2$ symmetry on $\phi$.
}
In this sense, the gravitational production studied in this paper provides a lower bound for the abundance of these particles.

\section{Discussion} \label{sec:dis}

We have seen that the gravitational particle production in supergravity can be significantly suppressed under some assumptions on the K\"ahler potential. It strongly depends on the choice of the K\"ahler potential of the chiral superfield as well as the inflaton sector.
However, it is remarkable that the effective gravitational interaction of a scalar field in supergravity can be drastically different from the non-SUSY case despite the minimal choice of the K\"ahler potential.

Finally we discuss several phenomenological implications of gravitational particle production in supergravity.
One immediate possible implication is production of purely singlet dark matter which interacts only through gravity.
Either scalar or fermionic component of $\chi$ described by the superpotential (\ref{W_sugra}) is a candidate of dark matter.
Assuming that the slow contribution is negligible, which is justified if $m_\chi \gg H_{\rm end}$, the abundance is given by the sum of (\ref{nfast_sugra}) (with $\epsilon$ replaced by $c_X$ including the effect of non-minimal K\"ahler potential) and (\ref{npsi}). In terms of the energy density to the entropy density ratio, we have
\begin{align}
	\frac{\rho_{\chi}}{s}&\simeq \frac{\mathcal C m_\chi H_{\rm end} T_{\rm R}}{4 M_P^2} \Delta \nonumber \\
	&\simeq 3\times 10^{-10}\,{\rm GeV}\,\Delta\,
	\left( \frac{\mathcal C}{10^{-2}} \right)
	\left( \frac{m_\chi}{10^{10}\,{\rm GeV}} \right)
	\left( \frac{H_{\rm end}}{10^{10}\,{\rm GeV}} \right)
	\left( \frac{T_{\rm R}}{10^{10}\,{\rm GeV}} \right),
	\label{rhochis}
\end{align}
where $T_{\rm R}$ denotes the reheating temperature of the universe and $\Delta$ is given by
\begin{align}
	\Delta \sim {\rm max}\left[  c_X^2,~\frac{m_\chi^2}{m_\phi^2} \right].
\end{align}
Therefore it is possible to obtain a right amount of $\chi$ dark matter.\footnote{
	One also should take into account the contribution from gravitational scattering of MSSM particle in thermal bath that create $\chi$ particles~\cite{Garny:2015sjg,Tang:2016vch,Tang:2017hvq,Garny:2017kha}.
}

It may also have some implications on the cosmological Polonyi/moduli problem~\cite{Banks:1993en,deCarlos:1993wie}.
Usually the moduli problem stands for the huge contribution from the coherent oscillation of the moduli.
Let us suppose that the coherent oscillation is suppressed somehow e.g. through the adiabatic suppression~\cite{Linde:1996cx,Nakayama:2011wqa} or just the fine-tuning of the initial condition.
Still the gravitational production channel is open~\cite{Giudice:1999yt} and it may give a lower bound on the moduli abundance.
If the moduli are stabilized by SUSY potential, the abundance of moduli/modulino is given by (\ref{rhochis}).\footnote{
	Here we do not consider the mixing of modulino with the inflatino and the SUSY breaking field that would make the story more complicated~\cite{Kallosh:2000ve,Nilles:2001fg,Ema:2016oxl}.
}
If the moduli are stabilized through SUSY breaking effect, one should take the SUSY mass $m_\chi\to 0$ in evaluating the moduli number density. In such a case the fermion contribution is negligible and the slow contribution to the scalar production should be taken into account. Then the moduli abundance is expected to be given by (\ref{rhochis}) after $m_\chi$ is reinterpreted as the SUSY breaking moduli mass and $\Delta$ replaced by
\begin{align}
	\Delta \sim {\rm max}\left[  c_X^2,~1 \right].
\end{align}
Compared with the upper bound on the abundance of hadronically decaying long-lived particle $\rho/s \lesssim 10^{-14}\,{\rm GeV}$ with lifetime longer than $10^2\,{\rm sec}$~\cite{Kawasaki:2017bqm}, one sees that the gravitational production of moduli can have impacts on cosmology.

\section*{Acknowledgments}

This work was supported by the Grant-in-Aid for Scientific Research C (No.18K03609 [KN]) and Innovative Areas (No.15H05888 [KN], No.17H06359 [KN]).



\end{document}